\documentclass[12pt]{article}

\usepackage{graphicx}
\usepackage[indention=0pt]{subfig}
\usepackage{amsmath}
\usepackage{amssymb}
\usepackage{xcolor}
\usepackage{times}
\usepackage[colorlinks=false, pdfborder={0 0 0}]{hyperref}
\newcommand{\superscript}[1]{\ensuremath{^{\textrm{#1}}}} 

\title{Observation of the Kibble-Zurek scaling law for defect formation in ion crystals}

\author
{S. Ulm,$^{1\ast}$ J. Ro{\ss}nagel,$^{1}$ G. Jacob,$^{1}$ C. Deg\"unther,$^{1}$ S.T. Dawkins,$^{1}$ U.G. Poschinger,$^{1}$ \\R. Nigmatullin,$^{2,3}$ A. Retzker,$^{4}$ M.B. Plenio,$^{2,3}$ F. Schmidt-Kaler,$^{1}$ K. Singer$^{1}$\\
\\
\\
\normalsize{$^{1}$QUANTUM, Institut f\"ur Physik, Universit\"at Mainz, Staudingerweg 7, 55128 Mainz, Germany}\\
\normalsize{$^{2}$Institut f\"ur Theoretische Physik \& Center for Integrated Quantum Science and Technology}\\
\normalsize{ Albert-Einstein-Allee 11, Ulm University, 89069 Ulm, Germany}\\
\normalsize{$^{3}$Department of Physics, Imperial College London, }\\
\normalsize{Prince Consort Road, London, SW7 2AZ, United Kingdom}\\
\normalsize{$^{4}$Racah Institute of Physics, The Hebrew University of Jerusalem, Jerusalem 91904, Givat Ram, Israel}\\
\normalsize{To whom correspondence should be addressed. E-mail:$^\ast$ulmst@uni-mainz.de.}\\
\\
\normalsize{Nature Communications 4, 2290 (2013), doi:10.1038/ncomms3290}\\
}

\date{\small{ }}

\begin{document}

\maketitle

\newpage

{\bf
Traversal of a symmetry-breaking phase transition at finite rate can lead to causally-separated
regions with incompatible symmetries and the formation of defects at their boundaries, which plays a
crucial role in quantum and statistical mechanics, cosmology and  condensed matter physics.
This mechanism is conjectured to follow universal scaling laws prescribed by the Kibble-Zurek mechanism.
Here, we determine the scaling law for defect formation in a crystal of 16 laser-cooled trapped ions,
which are conducive to the precise control of structural phases and the detection of defects. The experiment reveals an exponential scaling of defect formation $\gamma^\beta$, where $\gamma$ is the rate of traversal of the critical point and ${\bf \beta=2.68 \pm 0.06}$. This supports the prediction of ${\bf \beta = 8/3 \approx 2.67}$ for finite inhomogeneous systems. Our result
demonstrates that the scaling laws also apply in the mesoscopic regime and emphasises the potential
for further tests of non-equilibrium thermodynamics with ion crystals.
}
\vspace{1cm}


The Kibble-Zurek mechanism (KZM) applies to non-equilibrium systems traversing a second-order phase transition.
Prior to the phase transition, fluctuations relax into the lowest energy equilibrium state under the dissipative influence of a cooling mechanism.
The characteristic time of this relaxation increases near a structural phase transition and diverges at the critical point. Therefore, there is no finite rate at which the critical point can be traversed adiabatically and the structure of the system is effectively frozen before it reaches the critical point. Furthermore, different regions of the system can be causally disconnected due to a finite propagation speed of perturbations, allowing for multiple nucleation sites of the symmetry-broken ground states. If the choice of symmetries of neighbouring sections is incompatible, defects form where the phase boundaries meet and the system is thus prevented from reaching a global ground state.

The KZM was first proposed by T.W.B.~Kibble~\cite{kibble1976topology,kibble1980some} to describe the occurrence of topological defects in the early universe. Later W.H.~Zurek generalized this theory to condensed matter physics~\cite{zurek1985cosmological},
making this mechanism accessible in laboratories.
Experimental work on liquid crystals~\cite{chuang1991cosmology} and liquid $^4$He~\cite{hendry1994generation,ruutu1996vortex,bauerle1996laboratory} has provided confirmation of the KZM in a homogeneous setting.
Recent theoretical studies~\cite{PhysRevB.77.064111,Retzker2008DoubleWell,del2010structural,de2010spontaneous,Zurek2009} have triggered further interest in these scaling laws in other settings, such as inhomogeneous ion crystals confined in a linear Paul trap.

For such an inhomogeneous system, different regions reach the critical point at different moments and lead to time-dependent phase boundaries. This gives rise to an adiabatic transition if the propagation speed of perturbations $v_\textrm{s}$ exceeds the propagation speed of the phase boundary $v_\textrm{p}$, thus allowing for the causal connection of different domains and preventing the formation of defects. Hence, defects may only be created in a region of the system where  $v_\textrm{p} > v_\textrm{s}$.

Laser-cooled Coulomb crystals in ion traps feature the possibility of tunable spatial inhomogeneities.
In addition, this finite-sized mesoscopic system is accessible under highly-controlled experimental conditions, which permits the observation of the KZM without initial defects \cite{Griffin_PRX} or impurities.
Moreover, our setup features relevant control parameters that can be tuned over a wide range, ideal for the study of non-equilibrium statistical mechanics.

In this work, we investigate the KZM in the structural phase transition of an ion crystal from the linear to the zig-zag configuration \cite{PhysRevLett.85.2466,kaufmann2012precise}. The structure of localised ions in a linear Paul trap is controlled by the confining electrostatic potentials along the length of the crystal (axial) and the dynamic radio-frequency potential in the perpendicular (radial) directions, which gives rise to an effective harmonic potential in all three dimensions. The combination of the Coulomb repulsion between ions and the dissipative forces of laser-cooling yields a crystalline structure with minimum potential energy. For low axial trapping frequencies, this structure is a linear chain.
When the strength of the axial confinement is increased beyond a critical value, the structure is squeezed into the radial dimensions and undergoes a structural phase transition. In the case of a radially-anisotropic trap potential, the linear crystal transforms into a planar zig-zag configuration, which can be described by a second-order phase transition~\cite{PhysRevB.77.064111}. The latter phase can take the form of two symmetry-broken ground states, referred to as "zig-zag" and "zag-zig" configurations. Here, we generate a harmonic axial potential such that the phase transition initiates at the centre where the ion density is the highest.
The phase boundary then propagates outwards towards the ends with a finite propagation speed $v_\textrm{p}$.
If the curvature of the trapping potential is rapidly increased, $v_\textrm{p}$ can exceed the speed of sound $v_\textrm{s}$, and defects  may be created \cite{PhysRevLett.104.043004,0034-4885-75-2-024401}. Defects manifest themselves as kinks in the zig-zag structure of the crystal.
The positions of the individual ions are determined by fluorescence imaging, revealing the final configuration of the crystal.
This allows for the determination of the defect formation rate $d$ as a function of the rate at which the critical point is traversed.
By solving the time-dependent Ginzburg-Landau equation in the underdamped regime \cite{laguna1998critical}, a universal power law for the density of defects with respect to the rate of change of the control parameter, which is realised by the derivative of the trap frequency at the critical point $\gamma := d\omega/dt|_{\textrm{CP}}$, can be derived.
If the lengths of causally-connected regions are small compared to the system size, a scaling of the defect formation rate $d \propto \gamma^\beta$ with $\beta=4/3$ is predicted \cite{del2010structural,de2010spontaneous}.
Previous work has asserted that a further doubling of the exponent occurs~\cite{monaco2009spontaneous} when the system size is comparable to the size of these causally-connected regions, which leads to $\beta = 8/3\approx2.66$.

Our experiment with sixteen laser cooled ions is performed in this regime showing a scaling exponent of $\beta=2.68\pm0.06$ and $\beta=2.62 \pm 0.15$ for two different radial anisotropies, confirming the theoretical prediction. Our findings are reproduced by numerical simulations.

\begin{figure}
\begin{center}
\includegraphics[width=120mm]{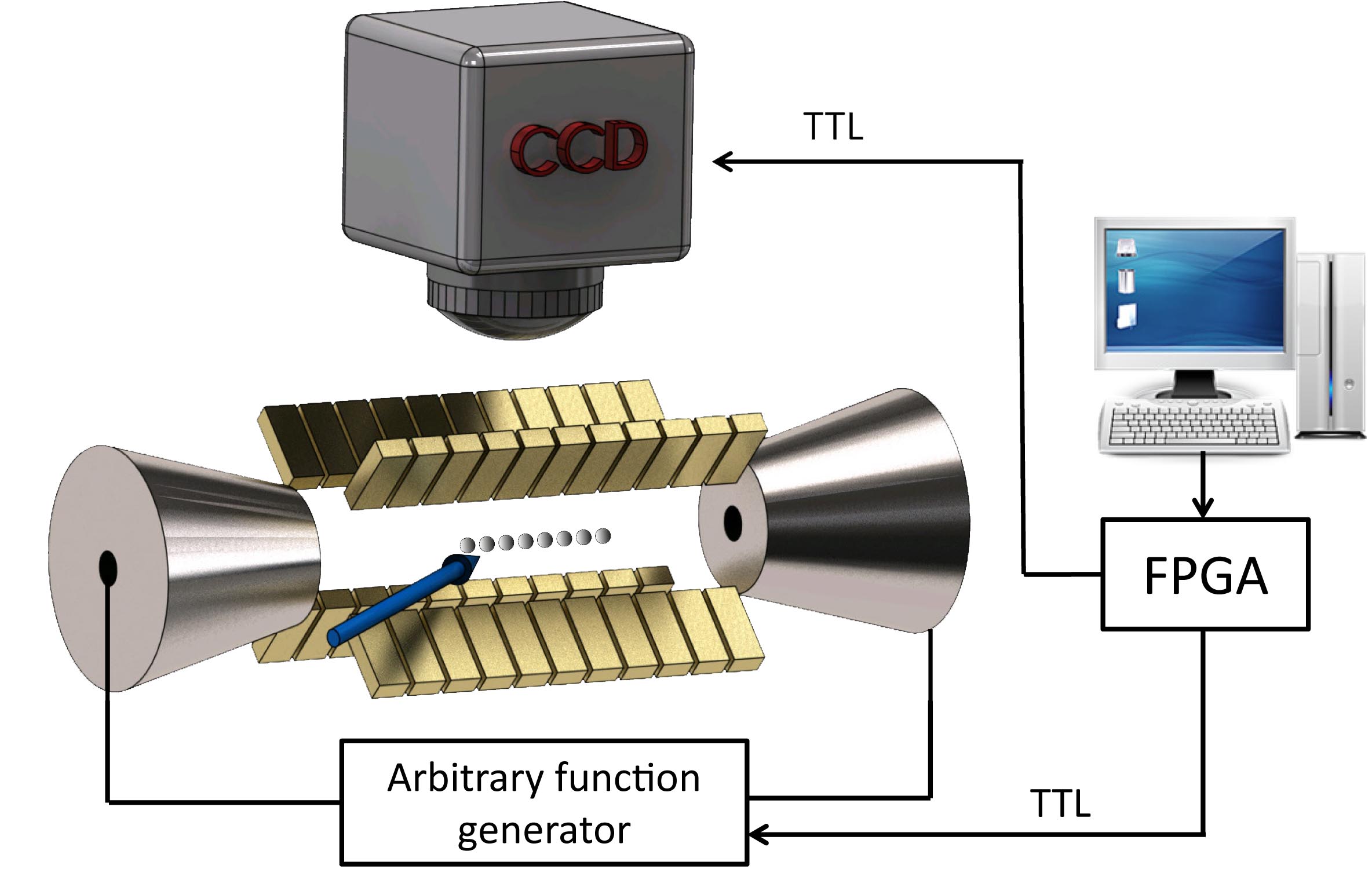}
\end{center}
\caption{
Schematic of the X-shaped segmented linear Paul trap. The cooling laser is oriented along the direction of the blue arrow. The end-cap voltages are controlled by an arbitrary function generator. The field-programmable-gate-array
(FPGA) controls the segment voltages and the triggering of the experimental sequence.
\label{fig:trap}}
\end{figure}

\begin{figure}
\centering
\includegraphics[width=85mm]{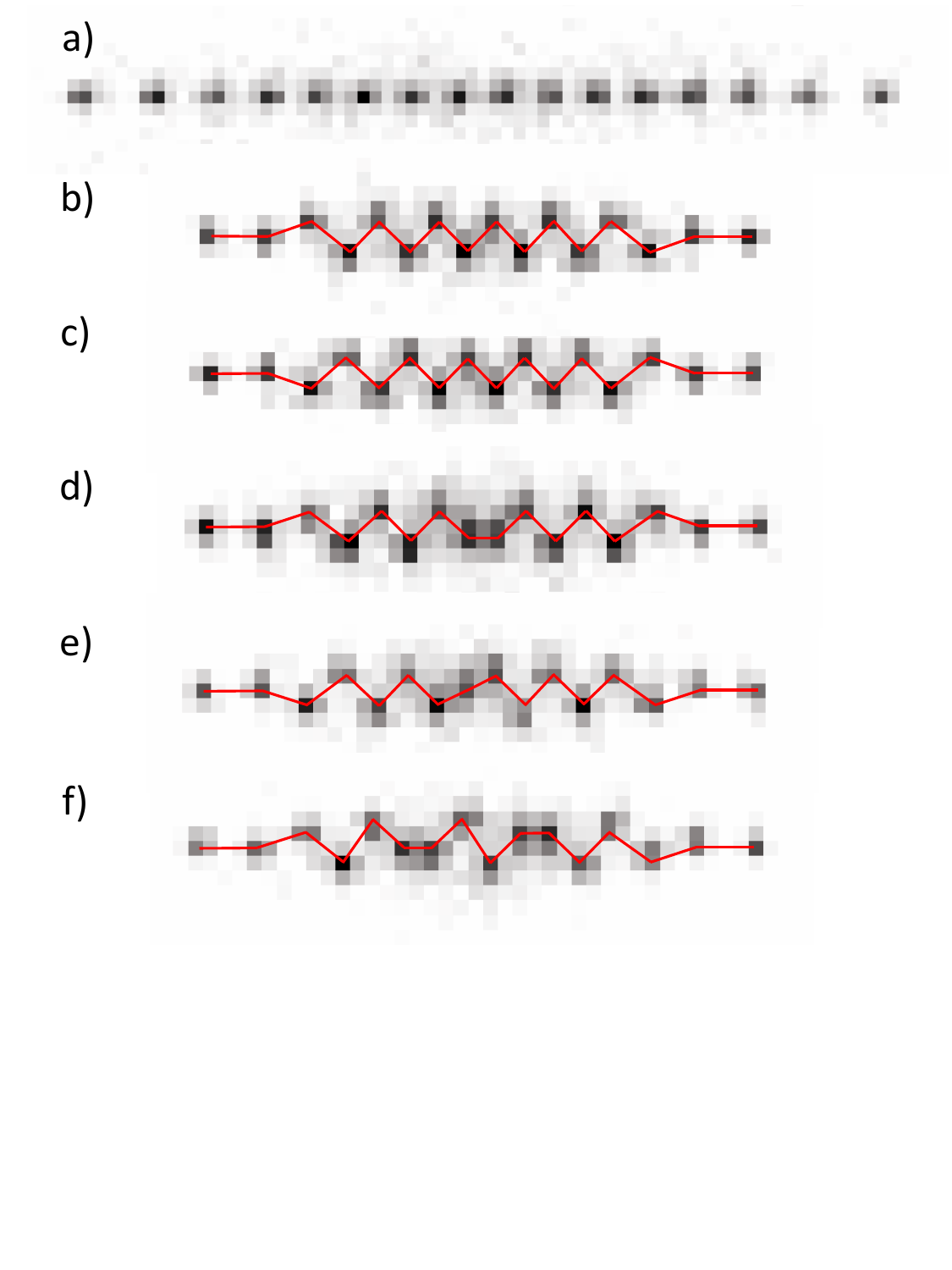}
\caption{Fluorescence images of a 16 ion crystal during the measurement of the KZM. a) Linear ion crystal before ramping the axial potential; b), c) zig-zag/zag-zig configuration after the ramp; d, e) appearance of single defects, which connect incompatible orientations of the crystal; f) double defects within the crystalline structure. The red line clarifies the configuration of the crystals. The width of one pixel corresponds to $1.7\,\mu\textrm{m}$.}
\label{fig:kinkology}
\end{figure}

\begin{figure}
\begin{center}
\includegraphics[width=90mm]{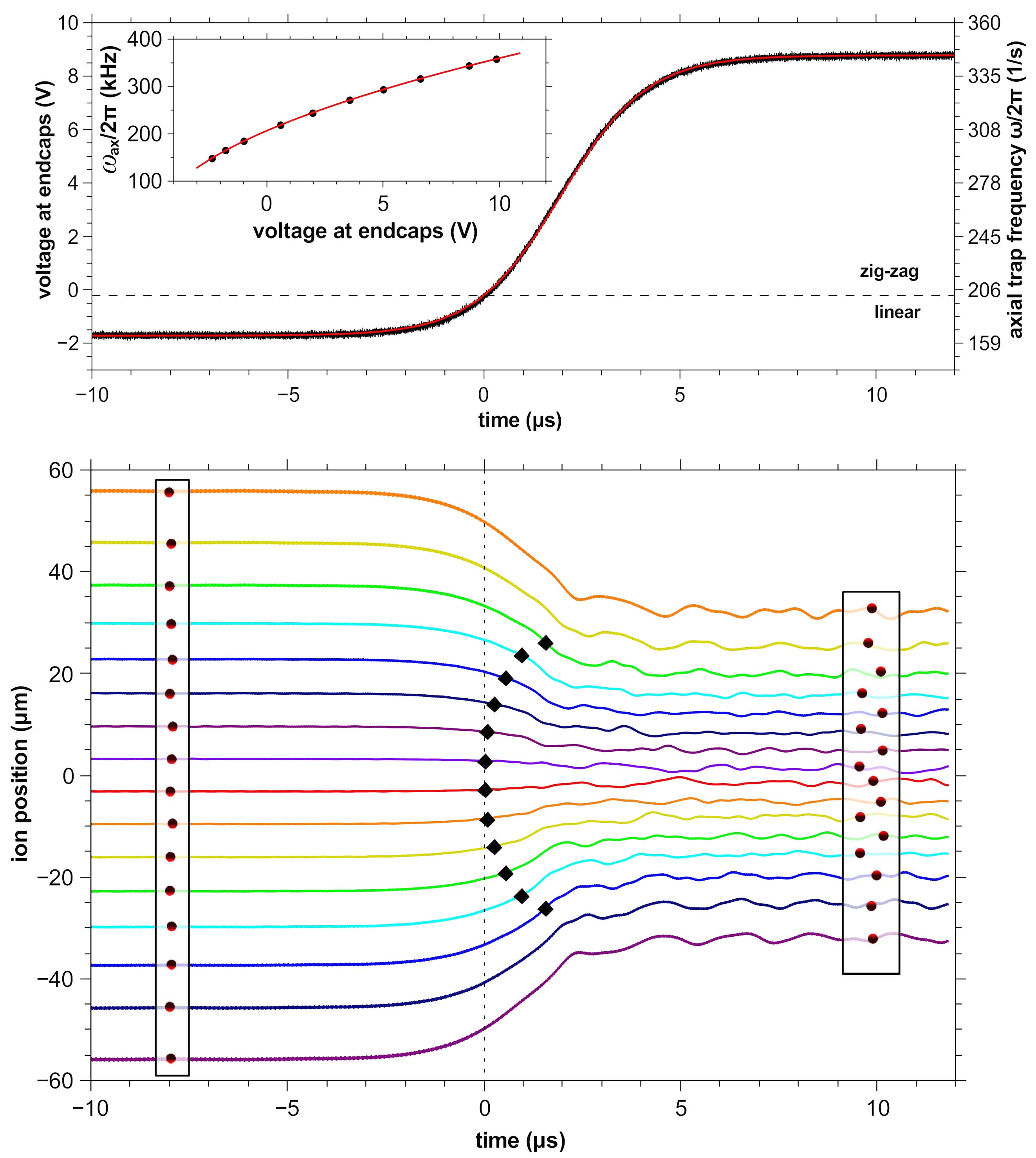}
\end{center}
\caption{Voltage ramp and corresponding crystal configuration.
Top: Measured voltage (left y-axis) applied to the end-caps by the arbitrary function generator (black line). A rounded shape of the waveform is chosen (red line) to avoid excessive excitation of axial vibrations. The timescale parameter $\tau$ determines the rate of change of the control parameter $\gamma$ at the critical point. The dashed line shows the separation between the two structural phases at an axial trap frequency of $\omega_{\textrm{ax}}/(2\pi)=201.7$~kHz and a radial trap frequency of $\omega_{\textrm{rad}}/(2\pi)=1394.1$~kHz. Inset: Dependency of the trap frequency $\omega_{\textrm{ax}}/(2\pi)$ on the applied end-cap voltage. A square-root function fits the measured data (red line). From this measurement, the functional dependency of the trap frequency on time and thus the rate of change of the axial frequency at the critical point is deduced, by  combining the functional dependence of the axial trap frequency as a function of the voltage at the endcaps with the measured ramp curve. The corresponding trap frequency can be obtained from the right axis. The scaling is obtained by combining the functional dependence of the axial trap frequency as a function of the voltage at the endcaps. Bottom: Axial positions of the 16 ions during the ramp as extracted from simulation results. Diamonds indicate the onset, if any, of the local phase transition for each ion, which are reached at different times due to the inhomogeneous charge density. The dashed line indicates the time when the middle ions reach the critical point. The corresponding crystal configuration is shown before and after the ramp. \\
\label{fig:experiment}
}
\end{figure}

\paragraph{Results}
\subparagraph{Scaling law}
In order to determine the scaling law of the defect formation rate, we load sixteen $^{40}$Ca ions into a linear segmented Paul trap (see Fig.~\ref{fig:trap}) realizing a linear ion crystal (Fig.~\ref{fig:kinkology} a). Doppler cooling close to saturation is applied during the whole experimental cycle. We ramp the axial trap frequency from $167\,$kHz to $344\,$kHz (see Fig.~\ref{fig:experiment} and Methods) across the critical point at $201.7\,$kHz, which leads to a change of the crystal structure into a "zig-zag" (Fig.~\ref{fig:kinkology} b) or "zag-zig" (Fig.~\ref{fig:kinkology} c) configuration. The voltage ramp is driven with an arbitrary function generator (AFG), which we used to generate an output with the functional form $V(t)\propto \left[1+\text{exp}(-(t-t_0)/\tau))\right]^{-1}$. 

\begin{figure}
\centering
\includegraphics[width=150mm]{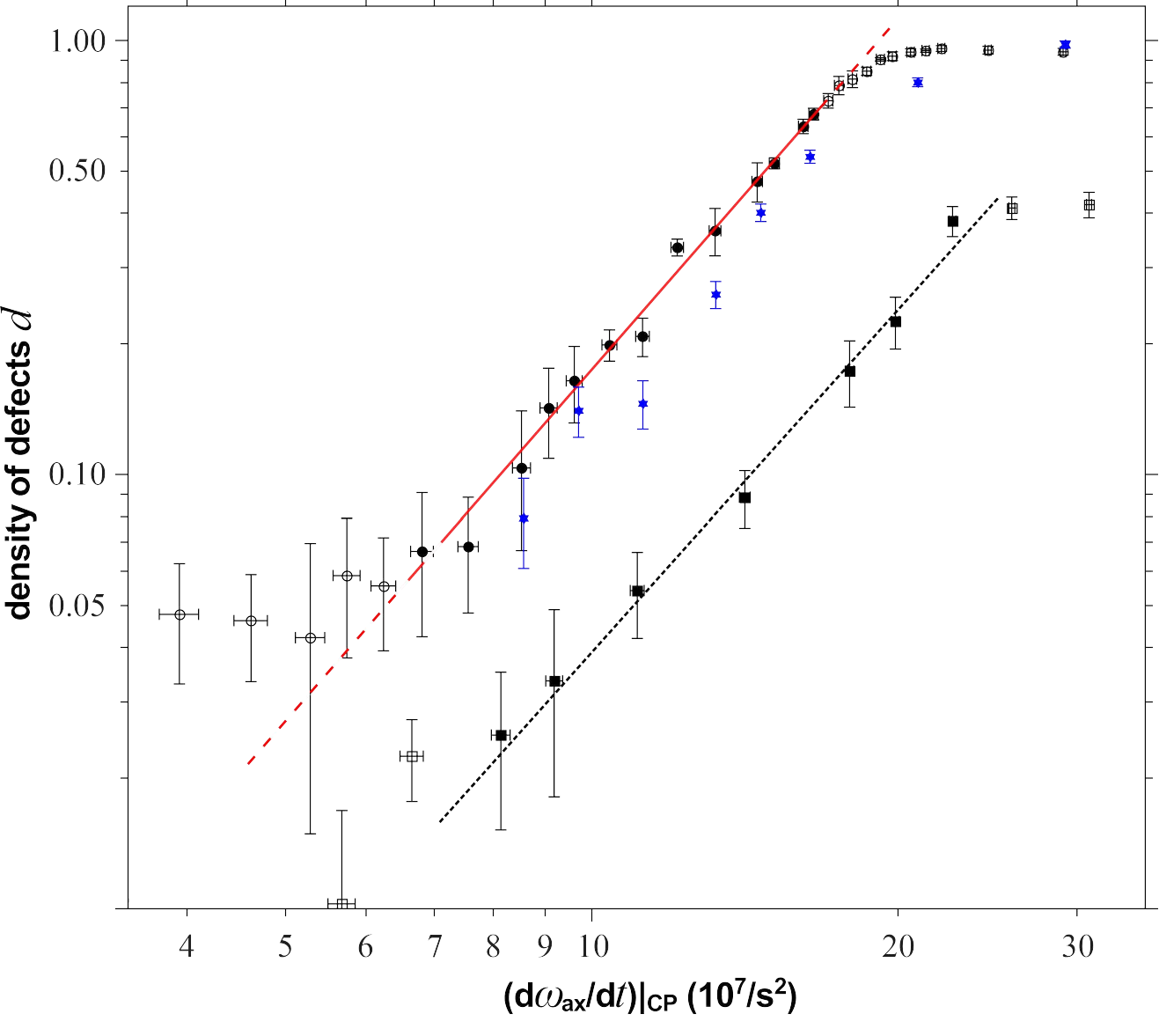}
\caption{Defect formation rates. Double-logarithmic plot of the measured density of defects $d$ versus the rate of change $\gamma$ of the axial trap frequency at the critical point. All circles correspond to 60000 measurements at a trap anisotropy of 1.03. The fitted function of the form $d \propto \gamma^{\beta}$ (red line) gives an exponent of $\beta=2.68\pm0.06$, which is in excellent agreement with the prediction of $\beta=8/3\approx2.67$.
The constant offset visible at lower ramping rates stems from background gas collisions at a base pressure of $1\times 10^{-9}$~mbar. This  is supported by the observation that the background rate  is directly proportional to the measurement time. The saturation is due to the maximum number of defects being limited by the system size.
For comparison, the rate of defects measured with a higher trap anisotropy of 1.05 is plotted (squares), showing ~50\% loss of defects but a similar fitted exponent of $\beta=2.62 \pm 0.15$ (dotted line).
Solid data points are used for the fits. The uncertainty in $\gamma$ is deduced from the scatter of repeated recordings of voltage ramps, while the uncertainty in $d$ is the standard deviation of the measurements. The blue stars depict the result of molecular dynamics simulations, see Methods for details.
}
\label{fig:results}
\end{figure}

This voltage was created at different rates with time constants $\tau$ ranging from $0.5\,\mu$s to $4.0\,\mu$s.
An image of the crystal configuration is captured $100\,\mu$s after the ramp with a $10\,$ms exposure time and the number of kinks (Fig.~\ref{fig:kinkology}d-f) is counted via image analysis (see Methods). The measured defect formation rate as a function of the rate of change of the axial trapping frequency $\gamma=\left( d\omega_{\textrm{ax}} / dt\right) |_{\textrm{CP}}$ is presented in Fig.~\ref{fig:results} (dots). The predicted scaling law for the inhomogeneous finite-sized system is confirmed by the observation of a scaling exponent of $\beta=2.68\pm0.06$, matching the prediction of $\beta = 8/3 \approx 2.67$.

\subparagraph{Effect of radial anisotropy}
Due to the anisotropy of the radial potential, the zig-zag phase is a planar crystal oriented in the plane of the weaker radial confinement. As shown in Fig.~\ref{fig:results} we have measured the defect density for two different radial anisotropies. With an anisotropy of $\omega_\textrm{x}/\omega_\textrm{y} = 1.03$, the observed defect densities are depicted as circles in fig.~\ref{fig:results}. For a small increase in anisotropy to $\omega_\textrm{x}/\omega_\textrm{y} = 1.05$, the defect density measured is decreased by approximately $50\,\%$ (see squares in Fig.~\ref{fig:results}), but the scaling observed, $\beta=2.62 \pm 0.15$ (Fig.~\ref{fig:results}) remains in agreement with the predicted value.

\paragraph{Discussion}

The influence of the crystalline structure on the dynamics and stability of defects can be described by the Peierls-Nabarro potential \cite{Peierls-Nabarro}. In our case, we have an attractive Peierls-Nabarro potential, with the defects trapped at the axial potential minimum, which is at the point of the highest charge density. This is based on the fact that we choose a low radial anisotropy such that, if defects form, they extend into the third dimension~\cite{mielenz2012trapping} and are likely to be axially-confined. For a small increase in anisotropy, this confinement is significantly reduced and thus the defect density is also diminished.  With a shallower Peierls-Nabarro potential, the thermal excitation of the ion crystal results in a higher loss rate of defects, because they migrate to the end of the crystal and vanish. This does not change the scaling behaviour because this loss channel acts as a multiplicative factor and thus shifts the curve in the vertical direction but does not influence the scaling exponent.

The experimentally determined scaling of defect formation is supported by molecular dynamics simulations of ion trajectories during a ramp (see Methods). A realistic model of the trap is employed and the equations of motion are solved with a partitioned Runge-Kutta integrator \cite{singer2010trapped,abah2012single}. Laser-cooling is modeled by a Langevin force and a constant dissipation.

The simulations reveal the dynamics of the ions on a fast timescale that is not accessible in the experiment and thus assist the experimental design and the interpretation of results. By monitoring the trajectory of each individual ion, we verified that no swapping of ions, and therefore no melting of the crystal, occurs during the ramping procedure. We also ensured that the excitation of axial oscillations is minimized (see Fig.~\ref{fig:experiment}) by choosing an optimized voltage ramp for squeezing the linear ion crystal over the phase transition. The simulation of the scaling behaviour is performed over the same range of critical parameters used in the experiment. The blue stars in Fig.~\ref{fig:results} show the simulation results with a scaling exponent of $\beta=2.53 \pm 0.23$ confirming the theoretical expectation. Note that the simulation result is shifted below the experimental data by around $5\,\%$, which corresponds to the defect generation rate due to background gas collisions.


In conclusion, we have observed the KZM in a model system with close-to ideal preparation, control and readout capabilities. The observed scaling of the defect formation rate for two different experimental conditions are in excellent agreement with the theoretical prediction. Realistic simulations show detailed agreement to experimental data also confirming the theoretical scaling exponent. In future experiments, we might utilise the trap control voltages to modify the local charge density and explore the role spatial inhomogeneities play in defect formation, thus investigating the crossover between the inhomogeneous and homogeneous KZM. Increasing the number of ions may allow access to the inhomogeneous KZM for large system sizes \cite{del2010structural}.
Instead of external trap potential ramps, spin dependent forces could be used to initiate structural phase transitions and quantum quenches \cite{PhysRevA.86.032104}. Ultimately, it might be possible to cool the system deeper into the quantum regime to explore quantum statistical mechanics where phase transitions are driven by quantum rather than thermal fluctuations \cite{damski2005simplest,zurek2005dynamics,Meyer2007,Shimshoni2011,Shimshoni2011A}. This would require all relevant vibrational modes of the linear ion crystal to be cooled near to the ground state, e.g. by employing electromagnetically induced transparency cooling \cite{Roos2000EIT}. In addition, by the use of sideband spectroscopy, defects in the zigzag crystal could be detected directly after the phase transition by their specific eigenmode, which is separated in the frequency domain from the modes of the regular crystal \cite{mielenz2012trapping}. During the preparation of this manuscript, we became aware of similar results presented in another paper \cite{pyka2012symmetry}.

\begin{figure}[h]
\centering
\includegraphics[width=85mm]{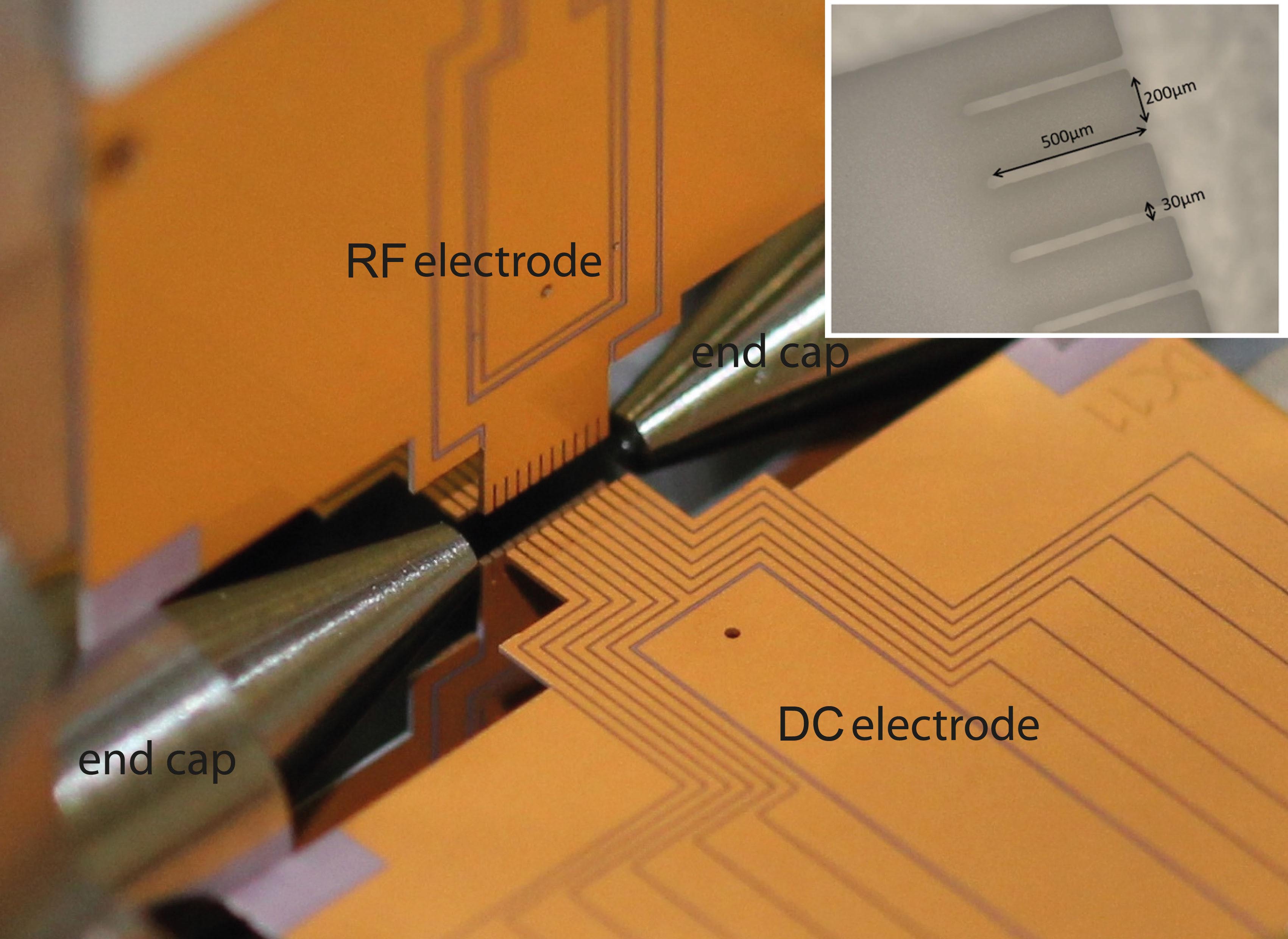}
\caption{Photograph of the segmented ion trap with gold coated alumina chips and conical endcaps. On the right hand side one sees a DC electrode with voltage supplies for the individual segments. On the top a RF electrode is shown, segmented for symmetry reasons. The angle of view corresponds to the direction of the cooling laser beam.}
\label{fig:trap2}
\end{figure}

\paragraph{Methods}
\subparagraph{Segmented linear Paul trap}
The experiments are performed in an X-shaped micro-fabricated segmented Paul trap based on four gold-coated laser-cut alumina chips each with 11 electrodes (see Fig.~\ref{fig:trap} and Fig. \ref{fig:trap2}) with a thickness of $125\,\mu$m. The segment width is $200\,\mu$m with isolating gaps of $30\,\mu$m. The radial distance of the segmented electrodes to the center of the trap measures $960\,\mu$m while the length of the whole trap is $2.9\,$mm. The radial confinement is generated by applying a radio-frequency voltage of approximately $450\,$V$_\textrm{pp}$ at a drive frequency of $\Omega/(2\pi)=22$\,MHz, resulting in a relevant radial trap frequency of $\omega_\textrm{rad}/(2\pi)$= $~1.4\,$MHz. The axial potential is generated by a superposition of static DC potentials applied to these segmented electrodes and variable voltages applied to the conical end-cap electrodes allowing for 

axial frequencies ranging from $\omega_\textrm{ax}/(2\pi)=167\,$kHz to $344\,$kHz. The base pressure in the vacuum chamber is $1\times 10^{-9}\,$mbar. Optical detection is achieved using an electron multiplying charge-coupled device camera (EMCCD camera) with a 10-ms-exposure time, oriented at $45^\circ$ to the planar structure of the crystal.

\begin{figure}[h]
\centering
\includegraphics[width=140mm]{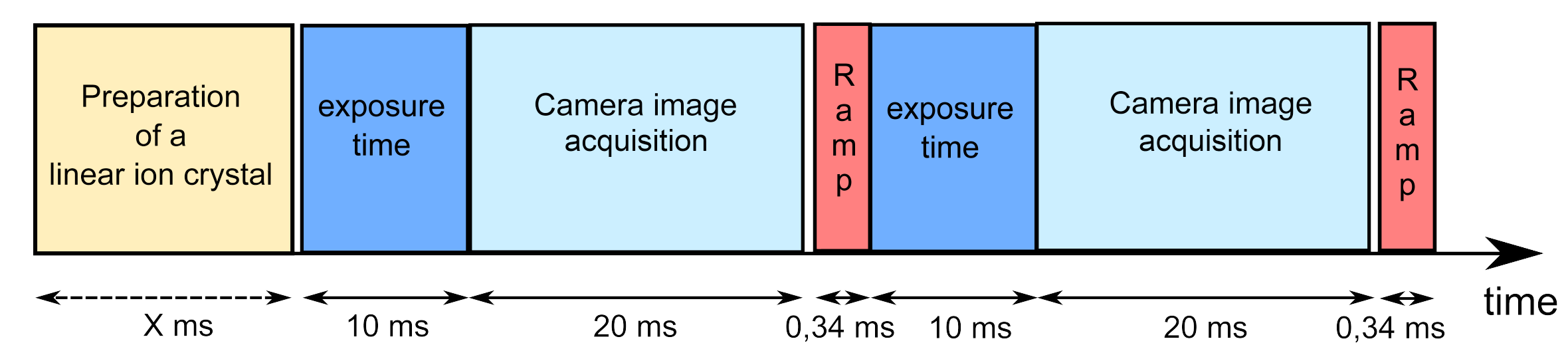}
\caption{Experimental sequence consisting of automatic preparation of a linear ion crystal consisting of 16 ions. Right before starting the ramping of the electrode voltages an image is taken with an exposure time of $10\,$ms. The image acquisition time amounts to $20\,$ms. Directly after the voltage ramp of typically 340$\,\mu$s another picture was taken. The experiment is then repeated by ramping down and conditional reloading of lost ions.}
\label{fig:sequence}
\end{figure}

The camera has an optical chip with $128\times 128$ pixels and a pixel size of $24\times 24\,\mu$m. An objective lens leads to an effective pixel size in the acquired images of $1.7\,\mu$m.
\subparagraph{Experimental sequence and setup}
In order to allow for high statistics and a large number of data points in maintainable time, a high repetition rate of the whole experimental sequence was ensured by implementing a fully automated real time experimental control system. An FPGA controls the timing of the lasers, the EMCCD camera and the AFG (see Fig.~\ref{fig:sequence}), which has 16 bit amplitude resolution at a sample rate of $200\,$MSamples per second for fast and glitch-free voltage ramps with variable time constants \cite{huber2008transport,walther2012controlling}. The experimental sequence (Fig.~\ref{fig:sequence}) starts with the loading of ions. For this purpose, Ca atoms in a thermal atom beam, cross the center of the trap and are ionized with resonantly enhanced two-photon ionization at wavelengths of $423\,$nm and $375\,$nm.

The cloud of trapped ions is Doppler cooled by laser radiation red detuned to the cycling dipole transition $^2S_{1/2} \leftrightarrow {^2P_{1/2}}$, leading to a condensation of the ions into a linear crystal. To drive the cooling transition we use a diode laser near $397\,$nm with a power of $0.15\,$mW and a beam direction which has a projection on all three trap axes. The cooling laser is switched on during the whole sequence of the experiment, and the resonance fluorescence is used for the image acquisition.

Ion losses can occur due to background gas collisions and possible losses during the relaxation ramp.
With the fully automated sequence we reach a repetition rate of one experiment per second:
In order to achieve automated control a first image is taken and the number of ions in the chain is evaluated. If the count is lower than 16, another pulse of the ionization lasers increases the number of ions. In the case that the number of ions is higher than 16, the axial trapping potential is temporarily lowered in order to reduce the ion number.
Only if the number of ions in the chain is exactly 16, does the sequence continue with the ramp of the axial confining potential. The FPGA then triggers the AFG which controls the ramp of the voltage at the endcaps and thus increases the axial trap frequency.

Due to noise on external signal filters, we observe a temporal jitter of the start of the voltage ramp of about $80\,\mu$s. To guarantee that the images are taken after the ramp, the image acquisition exposure of $10\,$ms starts with a delay of $100\,\mu$s. Subsequent to the ramp and the image acquisition, the axial potential is ramped back slowly to the initial value, such that the crystal relaxes again into a linear configuration. Another image is taken to verify the number of ions. If the number still equals 16, the ramping sequence is started again. Otherwise, the sequence continues with reloading.

\begin{figure}
\centering
\includegraphics[width=120mm]{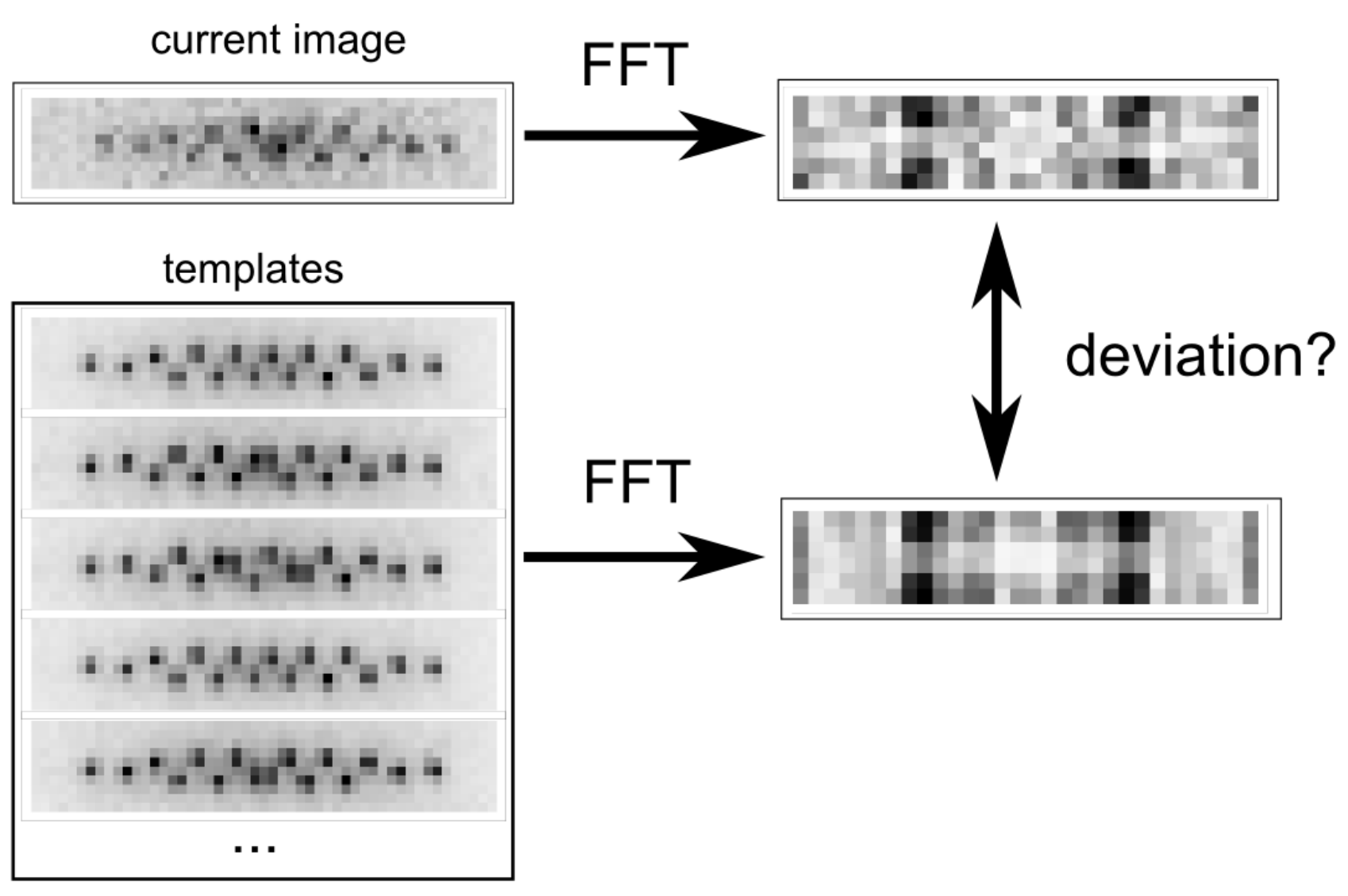}
\caption{Algorithm for image sorting. Image classification is performed by transforming the current image with the fast Fourier transformation (FFT) into the frequency domain. The normalized spectrograms are compared with normalized Fourier transformed template images and the $R^2$ deviation is used as a measure for classification.}
\label{fig:fft}
\end{figure}

\subparagraph{Evaluation of the defect density}To guarantee an efficient detection of defects and reliable categorisation of the images into different classes of possible crystal structures, their normalised two-dimensional discrete Fourier transforms are calculated and compared with 14 reference images (Fig.~\ref{fig:fft}). Such reference images were generated by averaging manually selected samples for a total of 14 relevant configurations. The coefficient of determination $R^2$ \cite{stapleton} is then used to classify the images into 15 classes corresponding to the 14 template images and one class for images which did not fulfill a matching threshold, containing blurred images and images where the amount of ions did not match 16. This recognition threshold is applied to the coefficients of determination to reject the low-quality images (less than 5$\%$). The density of defects is calculated as $d=(n_1+2\,n_2)/N$, where $n_1$ is the number of single-kink images, $n_2$ is the number of double-kink images and $N$ is the total number of images, not including those rejected.

\subparagraph{Long term Defect stability}In order to estimate the stability of the crystal structure during the fluorescence detection,
we repeatedly image the ions in the final potential. A blurred image indicates that the configuration changed during the exposure due to the loss of a defect or some other event, such as a background gas collision. For each of a range of exposure times from 10\,ms to 1800\,ms, more than 500 images were captured. The proportion of non-blurred images decreased exponentially with the
exposure time, revealing a time-constant of $\tau =(800\pm 140)\,\text{ms}$, the uncertainty being calculated assuming a binomial distribution for the two possible outcomes (blurred or non-blurred), combined with an error due to the sorting algorithm. This indicates that, in the 10\,ms required for imaging, the probability of a change to the crystal structure, including the loss of a defect, is less than a few percent.

\subparagraph{Molecular dynamics simulations}
In order to exclude the possibility of the crystal melting and vibrational excitation which would lead to a modified scaling behavior, we simulate the dynamics of the ions during the experiment using molecular dynamics (MD) Monte-Carlo simulations with a fourth order Runge-Kutta integrator performed on the bwGRiD computer cluster \cite{bwgrid}. In the simulations the system of ions is first initialized in a linear configuration at a specific temperature $T$, followed by the evaluation of the dynamics during the potential ramp.


To generate an ion chain at a specific temperature $T$, we used Langevin
thermostat MD calculation. The ions were assumed to obey the Langevin equation of motion

\begin{equation}
m\ddot{\textbf{r}}_{j}+\eta\dot{\textbf{r}}_{j}+\nabla_{j}V=\xi_{j}(t),\label{eq:supp1}
\end{equation}
where $\textbf{r}_{j}=(x_{j},y_{j},z_{j})$ is the coordinate of the $j$th
ion, $m$ is the mass of each ion, $V$ is the energy potential
of the ion chain, $\eta$ is the friction force and $\xi_{j}$ is
the stochastic force satisfying the following ensemble average relations

\begin{eqnarray}
\left\langle \xi_{j}(t)\right\rangle  & = & 0,\label{eq:supp2}\\
\left\langle \xi_{j}(t)\xi_{k}(t')\right\rangle  & = & 2\eta k_{\textrm{B}}T\delta_{jk}\delta(t-t').\label{eq:sup3}
\end{eqnarray}
The ponderomotive approximation for the potential energy is used:
\begin{equation}
V(t)=\frac{1}{2}\sum_{j=1}^{N}m\left(\omega_{x}^{2}x_{j}^{2}+\omega_{y}^{2}y_{j}^{2}+\omega_{z}^{2}z_{j}^{2}\right)+\sum_{j<k}^{N}\frac{1}{|\textbf{r}_{j}-\textbf{r}_{k}|}.\label{eq:supp4}
\end{equation}
Initializing the system of 16 Ca$^{+}$ ions with the starting conditions of the experiment results in a linear chain of ions at rest. Evolving the
system using equation (\ref{eq:supp1}) then leads to a chain
of ions at a specific temperature $T$, corresponding to the Doppler cooled ion chain in the experiment. Numerical evaluation was carried out using
the Langevin impulse method \cite{LI}. We have checked the thermalisation
of the system by checking that the equipartition theorem is satisfied
i.e. that $\left\langle \sum_{j}mv_{j}^{2}\right\rangle =3Nk_{\textrm{B}}T$,
where $v_{j}$ is the speed of the $j$th ion.


Once the system is thermalised, we evaluate the dynamics of the squeezed system by numerically solving the equations of motion with a time-dependent
trapping potential that matches the experiment. The electric potential, which depends on the electrode geometry, was modeled with a fast multipole boundary element electric field solver. Care was taken that experimental trap frequencies are reproduced.
The measured voltage ramps were fitted and the obtained values of the parameters were fed to the simulation program such that the simulated conditions are as close as possible to the experimental realization.
The motion of the system can be described by either Langevin equations or Newton's equations. We have checked that both Newton's equations and Langevin
equations lead to the same statistics of created defects. This is
due to the fact that during the quench the stochastic thermal fluctuations are
much smaller than the energy change due to the potential ramp. The simulations were carried out using a friction coefficient of $\eta=4.0\times10^{-21}\,$kg$\,$s$^{-1}$ and a Doppler temperature of $T=25\,$mK, corresponding to the experimental conditions. The integration step size was $2^{-20}/\omega_{\textrm{zi}}=0.895\,$ps with $\omega_{\textrm{zi}}$ the initial axial trap frequency. Around $3000$ simulated experiments are used for each data point.

\newpage
\paragraph{Acknowledgments.} The authors acknowledge helpful discussions with John Goold. The Mainz team acknowledges support by the Volkswagen-Stiftung, the DFG-Forschergruppe (FOR 1493) and the EU-project DIAMANT (FP7-ICT). MBP acknowledges support by the EU STREP project PICC (FP7-ICT), the Alexander von Humboldt Professorship and the GIF project "Non-linear dynamics in ultra-cold trapped ion crystals" and RN by the EPSRC Doctoral Training Center for Controlled Quantum Dynamics.




\end{document}